\newcommand\iso[2]{${\rm ^{#2}}$#1}
\def\aa{{A\&A}}

\def\aj{{AJ}}

\def\apj{{ApJ}}

\def\mnras{{MNRAS}}

\def\etal{\mbox{et al.}}

\def\deg{{$^{\circ}$}}
\def\bd17{BD +17\deg 3248}
\def\cs22{CS~22892$-$052}
\def\Msun{\mbox{$M_{\odot}$}}
\def\gtaprx{ \mathrel{  \vcenter{
                        \offinterlineskip \hbox{$>$}
                        \kern 0.3ex \hbox{$\sim$}    } } }
\def\ltaprx{ \mathrel{  \vcenter{
                        \offinterlineskip \hbox{$<$}
                        \kern 0.3ex \hbox{$\sim$}    } } }
\documentclass[cup5b]{caps}
\usepackage{graphicx}
\usepackage{amssymb}
\usepackage{ociwsymp4}   

\HeadText{J. J. Cowan and C. Sneden}

\begin{document}

\pagenumbering{arabic}

\author[]{JOHN J. COWAN$^{1}$ and CHRISTOPHER SNEDEN$^{2}$
\\
(1) Department of Physics and Astronomy, University of Oklahoma\\
(2) Department of Astronomy, University of Texas at Austin \\}

\chapter{Advances in $r$-Process \\ Nucleosynthesis}

\begin{abstract}
During the last several decades, there have been a number of
advances in understanding the rapid neutron-capture process
(i.e., the $r$-process).  These advances include large quantities of
high-resolution spectroscopic abundance data of neutron-capture elements,
improved astrophysical models, and increasingly more precise nuclear and
atomic physics data.  The elemental abundances of the heavy neutron-capture
elements,  from Ba through the third $r$-process peak, in low-metallicity
([Fe/H] $\ltaprx$ --2.5) Galactic halo stars are consistent with
the scaled (i.e., relative) solar system $r$-process abundance distribution.
These abundance comparisons suggest that for elements with Z $\ge$ 56
the $r$-process is robust---appearing to
operate in a relatively consistent
manner over the history of the Galaxy---and place stringent constraints
on $r$-process models.  While not yet identified, neutron-rich ejecta
outside of the core in a collapsing (Type II, Ib) supernova
continues to be a promising site for the $r$-process. Neutron star binary
mergers might also be a possible  alternative site.
Abundance comparisons of lighter $n$-capture elements in halo stars
show variations with the scaled solar $r$-process curve and might
suggest either multiple $r$-process sites, or, at least, different
synthesis conditions in the same astrophysical site.
Constraints on $r$-process models and clues to the progenitors of the halo 
stars---the earliest generations of Galactic stars---are also provided by 
the  star-to-star abundance scatter of [Eu/Fe]
at low metallicities in the early Galaxy.
Finally,
abundance observations of long-lived radioactive elements (such as Th and U)
produced in the $r$-process
can be used to determine the
chronometric ages of the oldest  stars,
placing constraints on the lower limit
age estimates of the Galaxy and  the Universe.
\end{abstract}

\section{Introduction}

Most of the heavy elements 
(here, Z $>$ 30) in the solar system are formed in
neutron-capture ({\it n}-capture) processes, either the slow ({\it s}-) or
rapid ({\it r}-) process.  
Our understanding of the distinction between these two processes follows
from the pioneering work of 
Cameron (1957) and Burbidge  et al. (1957). 
In the  $s$-process the relative lifetime for neutron captures 
($\tau_n$) is much longer than for electron ($\beta$) decays ($\tau_\beta$).
As a result,  the $n$-capture path in the $s$-process is near the 
so-called valley of beta stability, and the properties of nuclei involved
in this nucleosynthesis are, in great part, experimentally accessible.
The situation is quite different in the $r$-process where 
$\tau_n \ll \tau_\beta$. Thus, the $r$-process 
path occurs in a very neutron-rich
regime far from stability,   
making experimental measurements of those nuclei very difficult,  if not
impossible.

In this review we focus on advances in our understanding---still 
very incomplete---of the $r$-process. We note there have been a number of
earlier reviews, including those by Hillebrandt (1978), Mathews \& Cowan 
(1990), Cowan, Thielemann, \& Truran (1991a), Meyer (1994),  Truran et al. 
(2002), and Sneden \& Cowan (2003).  We employ and emphasize the 
observed stellar $n$-capture abundances in our discussion of the $r$-process.  
These abundances, particularly  in low-metal (i.e., low iron,
[Fe/H],  abundance) stars, 
provide direct  clues to the natures of
the $r$-process  and $s$-process formation sites.
In addition, the abundances of the $n$-capture elements  and 
related chemical evolution studies 
have also provided important information 
concerning  
the $r$-process, specifically  in relation to  
early Galactic nucleosynthesis and star formation history
(see reviews of Galactic chemical evolution by, e.g., 
Wheeler, Sneden, \&  Truran 1989, McWilliam 1997, and Truran et al. 2002).
We also note,  and discuss briefly,  the importance of certain
long-lived radioactive elements, 
such as thorium and uranium,   produced entirely in the {\it r}-process.
The abundance levels of these  nuclear chronometers
in the most metal-poor halo stars can provide  direct  age
determinations, 
and hence
set lower limits on Galactic and cosmological age estimates
(see, e.g., Cowan, Thielemann, \& Truran 1991b).

\section{Neutron-capture Abundances}

\begin{figure*}[t]
\includegraphics[width=1.00\columnwidth,angle=0,clip]{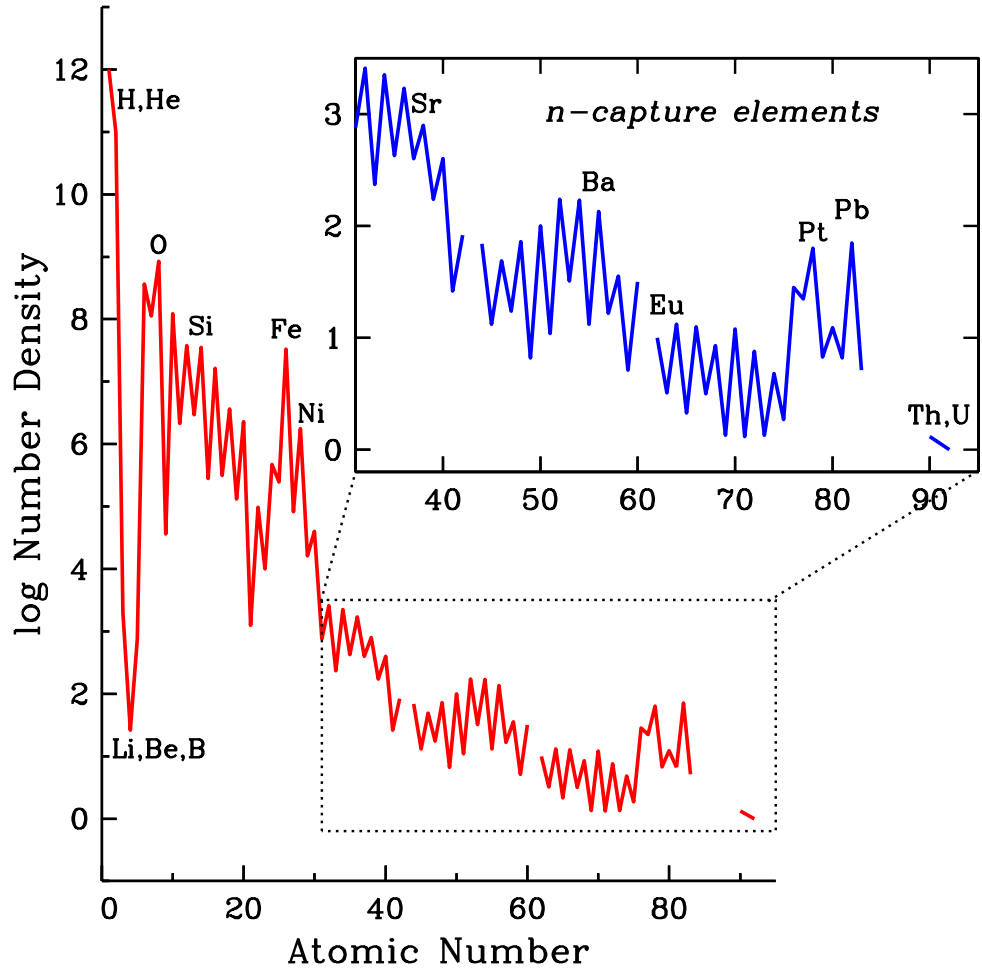}
\vskip 0pt \caption{
Abundances of elements in the Sun and in solar system material.
This abundance set is normalized by convention to log N(H) = 12.
The main figure shows the entire set of stable and long-lived radioactive
elements, while the inset is restricted to only those (neutron-capture) 
elements with Z~$>$ 30.
}
\label{fig1.ps}
\end{figure*}

In this section we examine the abundances of the elements, concentrating 
on those produced in neutron-capture processes.
We show in Figure~\ref{fig1.ps} 
the solar system abundances based upon the compilation  of
Grevesse \& Sauval (1998).  
Earlier compilations include those of 
Anders \&  Ebihara (1982), Cameron (1982a),
and Anders \&  Grevesse (1989).  
We also note the very recent solar system abundance determinations of
Lodders (2003). 
These solar system abundances can in many ways be treated as ``cosmic'' and
are frequently employed for stellar abundance comparisons.

\begin{figure*}[t]
\includegraphics[width=1.00\columnwidth,angle=0,clip]{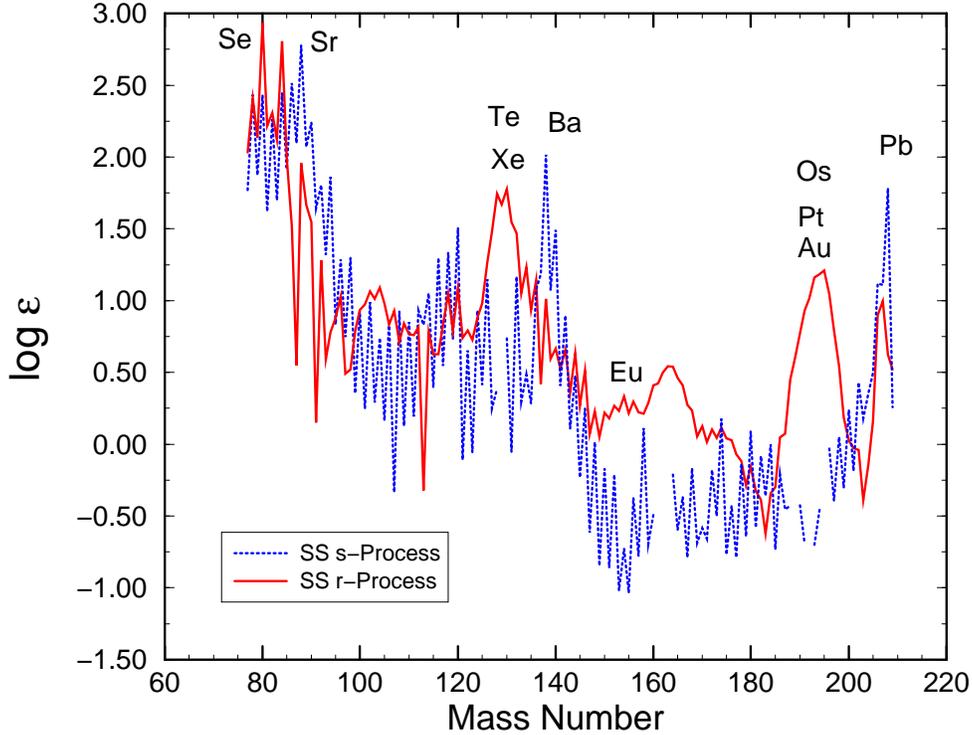}
\vskip 0pt \caption{
The $s$-process (dotted line) and $r$-process (solid line)
abundances in solar system matter,
based upon the work by K\"appeler et al. (1989). 
The total solar system abundances for the heavy
elements are from Anders \& Grevesse (1989).
}
\label{fig2.ps}
\end{figure*}

We highlight in  Figure~\ref{fig1.ps} the abundances of
the neutron-capture 
elements in solar system material.
As is well known, these elements above iron are synthesized predominantly by, 
and are the sum of 
individual isotopic contributions from, the {\it s}- and the
{\it r}-process. 
The deconvolution of the solar system material into the $s$-process and
$r$-process has traditionally relied 
upon reproducing the ``$\sigma$ N'' curve (i.e., the
product of the $n$-capture cross section and $s$-process abundance).
This ``classical approach'' to the $s$-process is
empirical and, by definition,  model independent. 
Subtracting these $s$-process isotopic contributions from the solar 
abundances  determines the residual $r$-process contributions.
Early deconvolutions  of solar system material into respective
$s$- and $r$-process contributions  were performed by Clayton et al. 
(1961),  Seeger, Fowler, \& Clayton (1965), and Cameron (1982b).
We show in Figure~\ref{fig2.ps}  a more recent such deconvolution based upon 
$n$-capture cross section measurements
(K{\"a}ppeler, Beer, \& Wisshak  1989; Wisshak, Voss, \&
K{\"a}ppeler 1996). In addition to this classical approach, 
more sophisticated
models, based upon $s$-process nucleosynthesis in low-mass AGB stars,
have been developed recently (Arlandini et al. 1999). 
Comparing the solar system elemental 
$r$-process abundance predictions obtained from these model calculations 
(Arlandini et al. 1999) with the classical approach 
(Burris et al. 2000) indicates  good agreement between each other
and with observed stellar abundances (Cowan et al. 2002;  
Sneden et al. 2003).  We also note in Figure~\ref{fig2.ps} 
that the $s$- and $r$-process solar system abundance distributions indicate that
not just individual isotopes but entire  elements were synthesized 
primarily in the $s$-process (e.g., Sr, Ba)
or the $r$-process (e.g., Eu, Pt) in solar system material.
These solar $s$-process, and corresponding $r$-process, elemental 
abundance distributions have been tabulated in, for example,
Sneden et al. (1996) and Burris et al. (2000). 

\section{Stellar Abundance Observations}
Many of the new advances in understanding the $r$-process have come
from stellar abundance observations, and we highlight some of those
critical new results in this section.

\subsection{Metal-poor Stars}

\begin{figure*}[t]
\includegraphics[width=1.00\columnwidth,angle=0,clip]{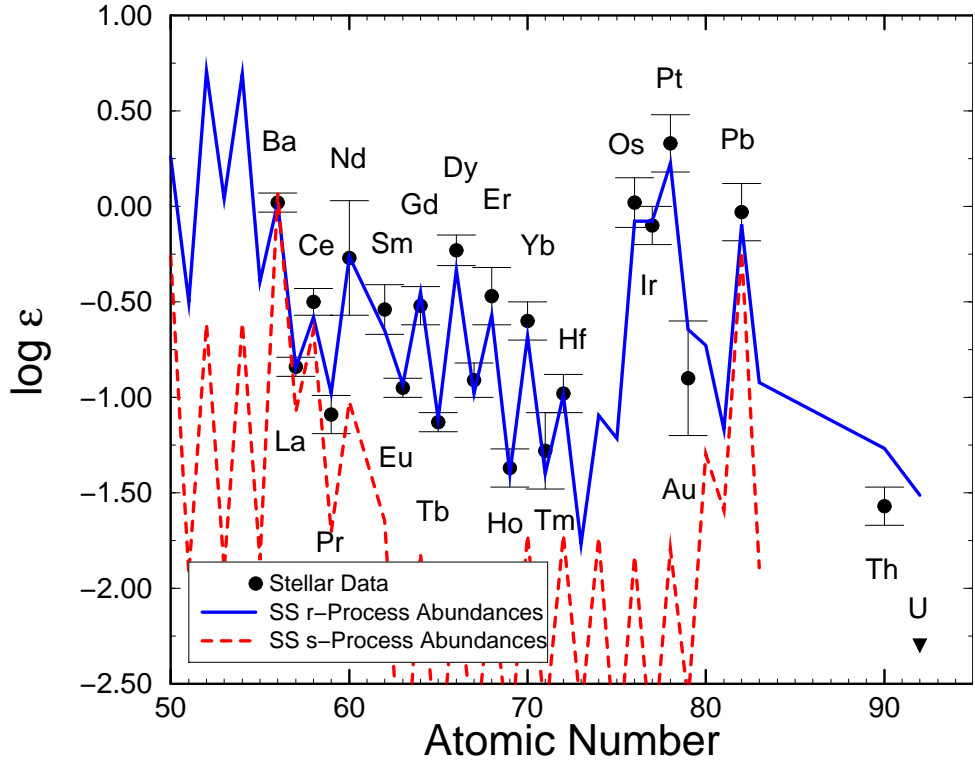}
\vskip 0pt \caption{
The heavy element abundance pattern for CS~22892$-$052, normalized to Ba, 
is compared with the scaled solar system $r$-process (solid line) and 
$s$-process (dashed line) abundance distributions.
}
\label{fig3.ps}
\end{figure*}

Various research groups over the last several decades have been employing the 
observed abundance distributions in metal-poor ([Fe/H] $<$ --1) Galactic halo
stars---bright giants with relatively ``uncrowded'' spectra---to try to 
identify, and to understand, the signatures of the $r$- and the $s$-process
(see, e.g., Spite \& Spite 1978; Sneden \& Parthasarathy 1983; Sneden \& 
Pilachowski 1985; Gilroy \etal\ 1988; Gratton \& Sneden 1994; Cowan \etal\ 
1995; McWilliam \etal\ 1995; Ryan, Norris, \& Beers 1996; Sneden \etal\ 1996;
Burris et al. 2000; Johnson \& Bolte 2001; Hill et al. 2002). 
These studies have all suggested  the dominance of the $r$-process in the 
oldest and most metal-poor Galactic halo stars. 
We show in Figure~\ref{fig3.ps} the abundances of $n$-capture elements
in \cs22 ([Fe/H] = --3.1) 
compared with scaled solar system $r$-process (Burris et al. 2000, solid line) 
and $s$-process (Burris et al. 2000, dashed line) 
abundance distributions (Sneden et al. 2003).
It is very clear that the $n$-capture element abundances in this star are
entirely consistent with the relative solar system $r$-process abundance
distribution. (The stellar abundances are also well fit with the solar system
$r$-process predictions of
Arlandini et al. 1999.) It is also clear that $s$-process nucleosynthesis
was not responsible for forming  the elements observed in this star,
at least in anything resembling solar proportions.

\begin{figure*}[t]
\includegraphics[width=1.00\columnwidth,angle=0,clip]{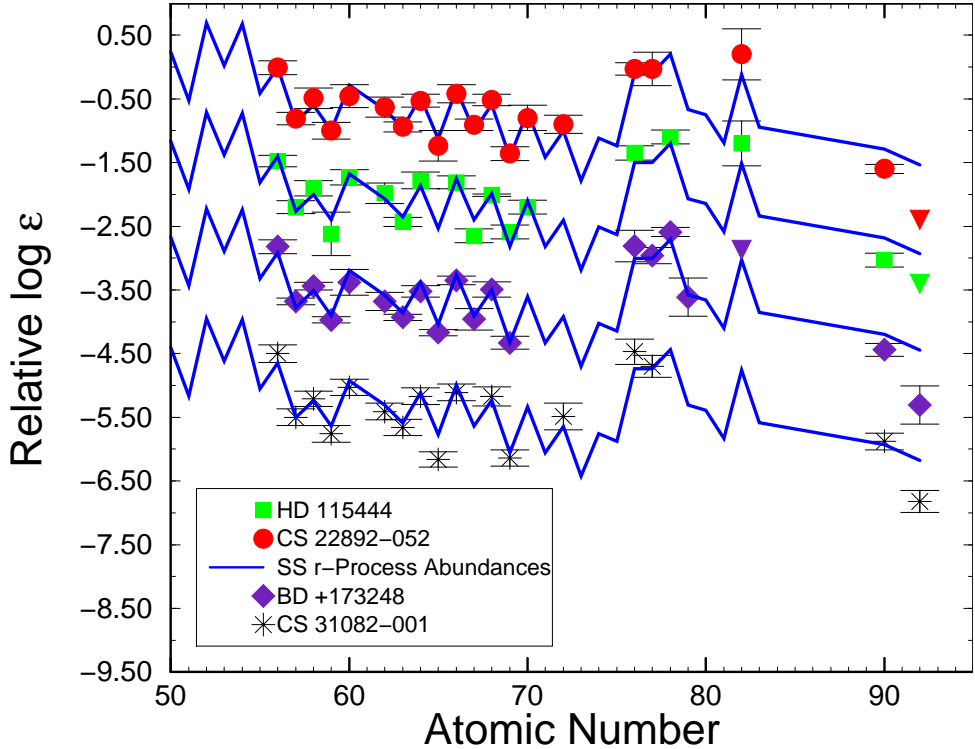}
\vskip 0pt \caption{
The  heavy element abundance patterns for the four stars CS~22892$-$052, 
HD~155444,  BD~+17$^\circ$3248, and CS~31082$-$001 are compared
with the scaled solar system $r$-process abundance distribution (solid line) 
(see Westin et al. 2000; Cowan et al. 2002; Hill et al. 2002; Sneden et al.
2003).  The absolute abundances have been shifted for all stars except 
CS~22892$-$052 for display purposes.  Upper limits are indicated by inverted 
triangles.  }
\label{fig4.ps}
\end{figure*}

Detailed abundance distributions,  with more than a 
few $n$-capture elements,  have been obtained for relatively few
cases---probably fewer than 20 of the metal-poor  halo 
stars. This picture has been changing in the last few years, however, with 
new comprehensive abundance 
studies  of the stars CS~22892$-$052 (Sneden 
et al. 1996, 2000a, 2003),  
HD~115444 (Westin et al. 2000), \bd17 (Cowan et al. 2002), and
CS~31082$-$001 (Hill et al. 2002). We show in 
Figure~\ref{fig4.ps} relative abundance distributions in those four stars, 
again compared with a scaled solar system $r$-process distribution (solid
lines).  Particularly noteworthy has been: (1) the increasing accessibility of
elements in the third $r$-process peak, typically only available with
observations in the UV with the {\it Hubble Space Telescope};
and (2) the increasingly precise abundance determinations,  resulting in great
part  from
marked improvements in the atomic physics data  
(see, e.g.,  Lawler, Bonvallet, \& Sneden  2001;
Den Hartog et al. 2003, and references therein). 
In fact, advancements in the atomic physics input have eliminated  more
and more discrepancies between 
observed metal-poor stellar abundances
and solar $r$-process abundance 
predictions.  Further improvements in individual elemental abundance
determinations might be employed to constrain  
the various theoretical predictions of the actual solar system $r$-process
abundances.

Figure~\ref{fig4.ps} 
makes clear that for all four of these $r$-process-rich stars,
the elemental abundances,  from Ba through the third $r$-process peak, 
are consistent with relative solar $r$-process proportions.
This suggests that for elements with Z $\ge$ 56 
the $r$-process is very robust,  appearing to
operate in a relatively consistent   
manner over the history of the Galaxy. This might imply a 
similar range of conditions (both astrophysical and nuclear) for the 
operation of the $r$-process (Freiburghaus et al. 1999a),
and  perhaps even a narrow range of masses for supernovae sites of
the $r$-process (e.g., Mathews, Bazan, \& Cowan  1992; Wheeler, Cowan, \& 
Hillenbrandt 1998; Ishimaru \& Wanajo 1999).

\subsection{Isotopic Abundances}

While there have been a continuing  number of  observations of elements in the 
metal-poor stars, isotopic abundance studies have been relatively rare. This
has been predominantly due to observational difficulties---isotopic 
wavelength shifts for transitions of most all
$n$-capture elements are small compared to the thermal and turbulent line
widths in stellar spectra. Recently, Sneden et al. (2002) 
and Aoki et al. (2003) 
have determined europium isotopic abundance fractions in  four 
very metal-poor, $r$-process-rich 
stars:  CS~22892$-$052, HD~115444, \bd17, and CS~31082$-$001.
The abundance fractions for Eu  in these stars 
are in excellent agreement with each other and with their values in
the solar system: fr(\iso{Eu}{151})~$\simeq$ fr(\iso{Eu}{153})~$\simeq$ 0.5.
Additional Eu stellar abundance studies that demonstrate this same solar system
$r$-process agreement  have been
reported (Ivans 2003, private communication).
These  isotopic abundance observations  support earlier studies that indicate
that stellar elemental 
abundances for Z~$\geq$~56 match very closely those of a scaled
solar system $r$-process abundance distribution.
With only two isotopes, perhaps it is not totally surprising that the Eu abundance
fractions in the metal-poor halo stars are the same as in the solar system.
It does suggest, though, that the solar abundances are cosmic,
in some sense, and that the $r$-process (for the heavier $n$-capture elements) 
is  robust over the lifetime of the Galaxy. 
Nevertheless, a more definitive test would be an isotopic analysis of 
the element Ba with many
more isotopes than Eu.
Lambert \& Allende-Prieto (2002) have done such an analysis in another metal-poor star,
HD 140283, and find that the isotopic fractions of
Ba are also consistent with the solar $r$-process values.
Additional stellar isotopic abundance studies will be necessary to 
strengthen and extend these findings.

\section{The $r$-Process }

Although the $r$-process has been studied for many years, the actual site for this
nucleosynthesis  has not been identified. Further complicating this  search 
is the possibility of more than one such site.
Nevertheless, there has been much progress in our understanding of the astrophysical
models and the related nuclear physics of the $r$-process. 

\subsection{Astrophysical Sites and Models}
The nature of the $r$-process requires high neutron number densities on short time scales,
indicative of explosive environments. The early work of Burbidge et al. (1957) 
suggested that the neutron-rich ejecta  outside of
the core in a collapsing (Type II, Ib) supernova 
was the likely site for the $r$-process. Nevertheless, the detailed physics of
core-collapse supernovae were poorly known at that time, to say nothing of
the lack of computational tools. These hindrances prevented definitive 
identifications on  the nature of the $r$-process and led to the  
consideration of other possible sites, including
the shocked helium and carbon zones of exploding 
supernovae and jets and bubbles of neutron-rich material ejected from the
collapsing core 
(see Cowan et al. 1991a, and references therein). 
Inhomogeneous Big Bang cosmological models
were even studied as possible sites (Rauscher et al. 1994). 

Advances in understanding 
supernova physics,  particularly neutrino interactions, led to new  promising 
$r$-process scenarios,
such as the high-entropy neutrino wind in supernovae
(Takahashi, Witti, \& Janka  1994; Woosley et al. 1994; 
Qian \& Woosley 1996; Wanajo et al. 2001, 2002; Terasawa et al. 2002).
There have been some problems, 
however, with these models in obtaining the required entropies and
in some inadequate abundance predictions 
(see, e.g., Meyer, McLaughlin, \& Fuller 1998;
Freiburghaus, Rosswog, \& Thielemann 1999b; Thompson, Burrows, \& Myer 2001; 
but see also Thompson 2003). 
(See Thielemann et al. 2002 for a general review of nucleosynthesis in 
supernovae and the related model uncertainties.)
While much emphasis has been placed on determining the physics in 
``delayed'' models, 
``prompt'' supernova explosion scenarios  have not been abandoned as a 
possible  site for the $r$-process (see, e.g., Wheeler et al.  1998; 
Sumiyoshi et al. 2001; Wanajo et al. 2003). 
It has also been argued that not all core-collapse supernovae are responsible
for $r$-process synthesis. In particular there have been a number of 
studies  that suggest only low-mass ($\ltaprx$) 11 \Msun supernovae are 
likely sites (Mathews \& Cowan 1990; Mathews et al. 1992; 
Wheeler et al. 1998; Ishimaru \& Wanajo 1999; Wanajo et al. 2003; 
but see also Wasserburg \& Qian 2000 or  Cameron 2001).

While most of the attention in studying the $r$-process has focused on 
supernovae, there  has been some consideration of  
neutron star binaries, which have  an 
abundance of neutron-rich material. 
Early studies suggested that  
the tidal interaction  between a
neutron star and a black hole,  or a second  neutron star, 
might be a possible astrophysical
site for this nucleosynthesis
(see, e.g., Lattimer et al. 1977). 
Despite  encouraging recent studies  by Rosswog et al. (1999) and 
Freiburghaus et al. (1999a), however,  
there are questions about whether 
the frequency of these events  and the amount of 
$r$-process ejecta per merger are consistent with observational constraints
(Qian 2000).

Accompanying the advances in these 
more sophisticated astrophysical models has been a
concomitant improvement  in our understanding of the nuclear physics involved 
in the $r$-process - particularly more reliable nuclear information about the
very neutron-rich nuclei.
The $r$-process occurs  far from stability, and, thus, in the past there has been little
reliable nuclear data available. 
Recently, however, there have been 
an increasing amount of experimental
determinations of critical nuclear data, including  
half-lives and neutron-pairing energies (see, e.g., Pfeiffer, Kratz, \&  
M\"oller 2002; M\"oller, Pfeiffer, \& Kratz 2003).
In addition to these new nuclear data, 
there have been recent advances in theoretical prescriptions for very
neutron-rich nuclear
data (see, e.g., Chen et al. 1995; Pearson, Nayak, \& Goriely
1996; M\"oller, Nix, \& Kratz 1997).
In particular, these developments  include  
nuclear mass formulae that are more reliable
and physically predictive for nuclei far from stability---especially crucial
for chronometer studies---for example,  
such mass models  as 
ETFSI-Q and HFBCS-1  (see Schatz et al. 2002 for discussion and additional references therein).
The combination of 
more nuclear
data and advances in theoretical mass models has 
led to increasingly more reliable
descriptions  for very
neutron-rich  nuclei, 
necessary for a better understanding of the $r$-process
(see also Pfeiffer et al. 2001 for further discussion) 

\subsection{Two $r$-Processes?}
\begin{figure*}[t]
\includegraphics[width=1.00\columnwidth,angle=0,clip]{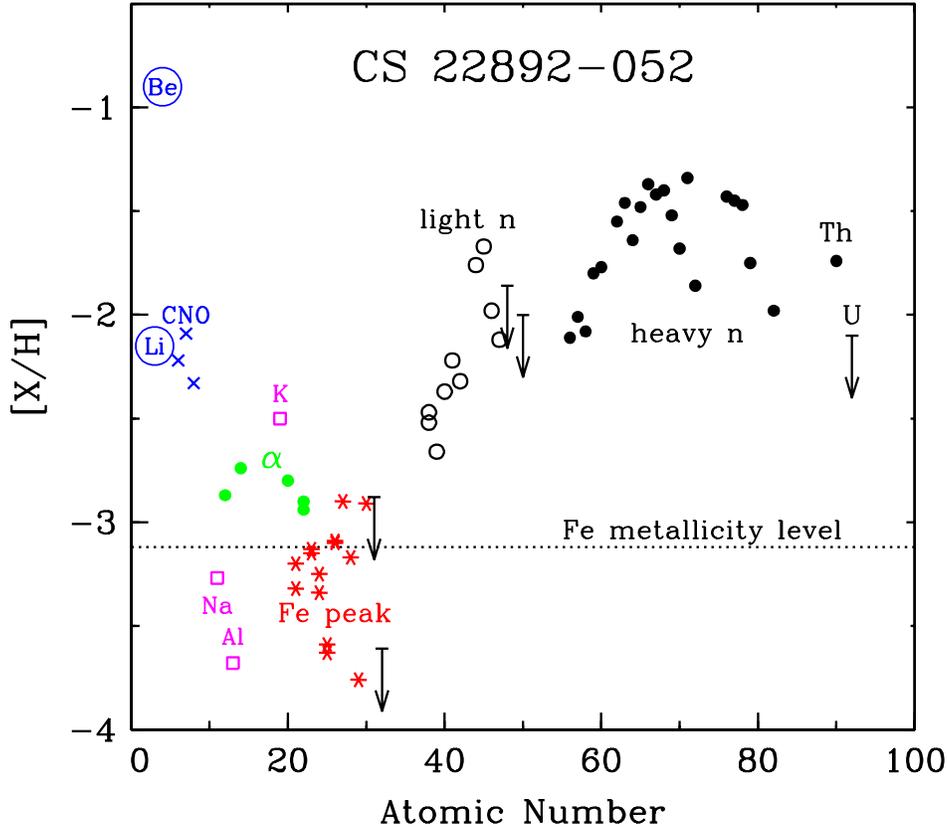}
\vskip 0pt \caption{
Total abundance pattern in \cs22 with respect to solar system 
values. The dashed line 
represents the iron abundance (i.e., the metallicity of the star). 
Upper limits are denoted by downward-pointing arrows. Li and Be are displaced
from their actual abundance values for display purposes.
}
\label{fig5.ps}
\end{figure*}

The observations (discussed above) 
demonstrate  that the
heavier (Ba and above, Z $\ge$ 56, or A $\gtaprx$ 130--140)
neutron-capture elements,  
particularly in $r$-process-rich stars,
are consistent with a scaled solar system $r$-process curve. 
Until very recently, however, there has been relatively little data  for 
elements between Zr and Ba.  
We show in Figure~\ref{fig5.ps} the total abundance summary of the 
elements in 
\cs22. A total of 58  elements (53 detections and 5 upper limits) 
have been observed in this star, 
which appears to be 
the most of any other star except the Sun at the time (Sneden et al. 2003).
The dashed line in the figure indicates the iron abundance 
([Fe/H] = --3.1) in this star.  
It is clear from the abundances of the heavy $n$-capture elements
why this star has been so well studied---[Eu/Fe], for example, is   enhanced 
by approximately a factor of 45 above the iron abundance level.  It is also 
seen in this figure that the abundances of the light $n$-capture elements in 
the little-explored element regime of 
Z = 40--50 mostly lie below those of the heavy $n$-capture elements.

\begin{figure*}[t]
\includegraphics[width=1.00\columnwidth,angle=0,clip]{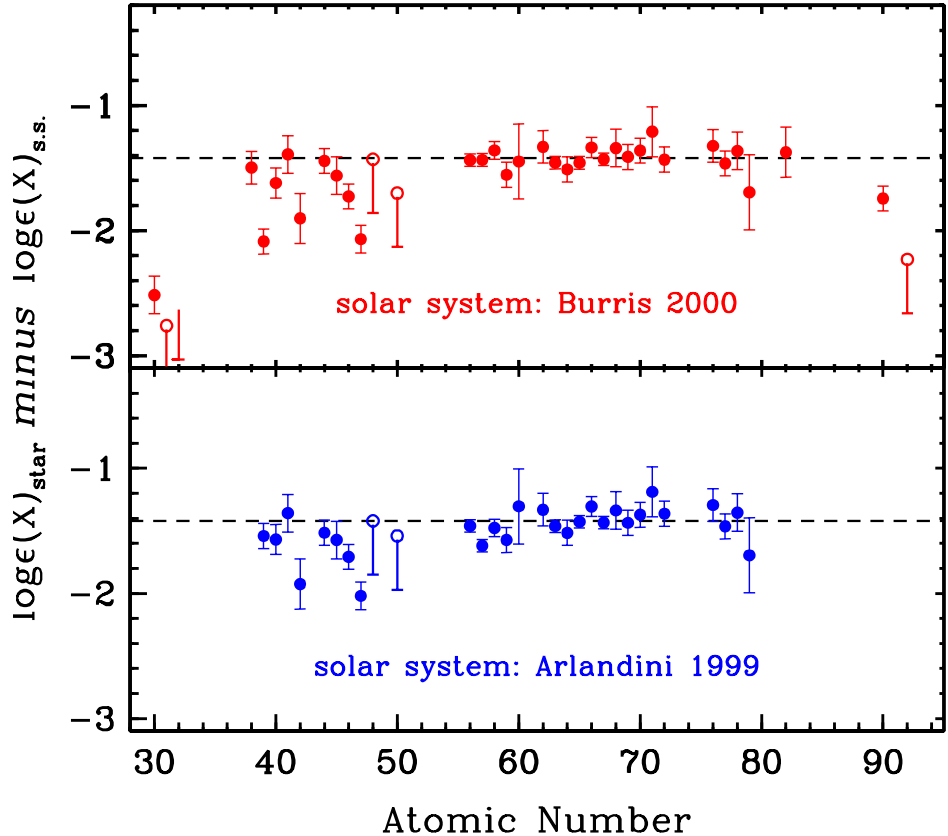}
\vskip 0pt \caption{
Differences between \cs22 abundances and two scaled solar system abundance
distributions, after Sneden et al. (2003).
The dotted line in each panel indicates the unweighted mean difference
for elements in the range 56~$\leq$~Z~$\leq$~79.
In the top panel the abundance differences relative to those of
Burris \etal\ (2000; their Table~5) are shown.
Upper limits are indicated by open circles, except for Ge 
(Z~=~32), because its upper
limit lies below the lower limit boundary of the plot.
In the bottom panel the abundance differences are relative to those
of Arlandini \etal\ (1999),
who tabulated $n$-capture abundances only for
the atomic number range 39~$\leq$~Z~$\leq$~81.}
\label{fig6.ps}
\end{figure*}

This difference is seen in more detail in Figure~\ref{fig6.ps}, where 
the abundances in \cs22 (Sneden et al. 2003) 
are compared with two predictions for solar system $r$-process abundances, by 
Burris et al. (2000; top panel) and Arlandini et al. (1999; bottom panel).
The dotted line in each panel indicates the unweighted mean difference
for elements in the range 56~$\leq$~Z~$\leq$~79.
It is clear that the stellar abundance 
data are well fit by both of these distributions
for Z$\ge$ 56,  confirming  earlier such results (as discussed above and
shown in  Figs.~\ref{fig3.ps} and \ref{fig4.ps}). 
The data, however,   
seem to indicate that some of the lighter $n$-capture
elements from Z = 40--50 
(for example Ag and Mo) are not consistent with (i.e., in general fall below) those
same scaled $r$-process curves that fit the heavy $n$-capture elements. 
There are exceptions, with the abundances 
of Nb and Rh seemingly  consistent with the scaled solar system $r$-process
curve, but 
on average these lighter elements do seem to have been
synthesized at a lower abundance level than the heavier $n$-capture elements.

There are several possible explanations for the differences in
the abundance data for the lighter and heavier {\it n}-capture elements.
These observations might 
support earlier suggestions of two $r$-processes based
upon solar system meteoritic (isotopic) data (Wasserburg, Busso \& Gallino 1996).
It has been suggested, for example, that perhaps, analogously to the
$s$-process, the lighter elements might be synthesized in a ``weak''
{\it r}-process with the heavier elements synthesized in a  more robust
``strong'' (or ``main'') {\it r}-process (Truran et al. 2002).
Thus, the helium zones of exploding supernovae, 
have been suggested as possible second {\it r}-process sites that might
be responsible for the synthesis of nuclei with A $\lesssim$ 130--140
(Truran \& Cowan 2000).
Or the two sites might come from 
supernovae of a different mass range or frequency (Wasserburg \& Qian 2000),
or perhaps a combination of supernova and neutron star
binaries (Truran et al. 2002).
Alternative interpretations have suggested  that the entire abundance
distribution could be synthesized in a single core-collapse supernova
(Sneden et al.  2000a; Cameron 2001).
We note,  however,  that only a few (three) 
stars have detailed abundance distributions that 
include data in this  Z = 40--50 element domain.  
Crawford et al.
(1998), however, detected silver in four halo stars,
and Sr, Pd, and Ag abundances
for a sample of metal-poor stars  
have been reported by Johnson \& Bolte (2002). 
Further detailed spectroscopic 
studies, in conjunction with additional theoretical efforts,
 will be necessary to determine any differences in the nature and
history of  
the synthesis of the lighter and heavier $r$-process elements.

\section{$r$-Process Abundance Scatter in the Galaxy}

A number of observational and theoretical studies have 
demonstrated that at the earliest times in the Galaxy the $r$-process was
primarily responsible for $n$-capture element formation, even for elements
(such as Ba) that are formed primarily in the $s$-process in solar system 
material
(Spite \& Spite 1978; Truran 1981;
Sneden \& Parthasarathy 1983;
Sneden \& Pilachowski 1985; Gilroy \etal\ 1988;
Gratton \& Sneden 1994; McWilliam \etal\ 1995;
Cowan \etal\ 1995; Sneden \etal\ 1996;
Ryan et al. 1996).
The presence of these $r$-process 
elements in the very oldest  stars  in our Galaxy
strongly  suggests the astrophysical $r$-process site is short-lived.
Thus, the first stars, the progenitors of the halo stars,  
were likely massive and evolved quickly, 
synthesized the $r$-process elements and ejected them
into the interstellar medium before the formation of the currently observed  stars.
In contrast,  
the primary site for $s$-process nucleosynthesis 
is 
low- or intermediate-mass stars (i.e., 
$M \simeq 0.8-8$ \Msun) with  long evolutionary time scales 
(Busso, Gallino, \& Wasserburg 1999).
Thus,  these stars  would not have had time to
have synthesized the first elements in the Galaxy.

\begin{figure*}[t]
\includegraphics[width=1.00\columnwidth,angle=0,clip]{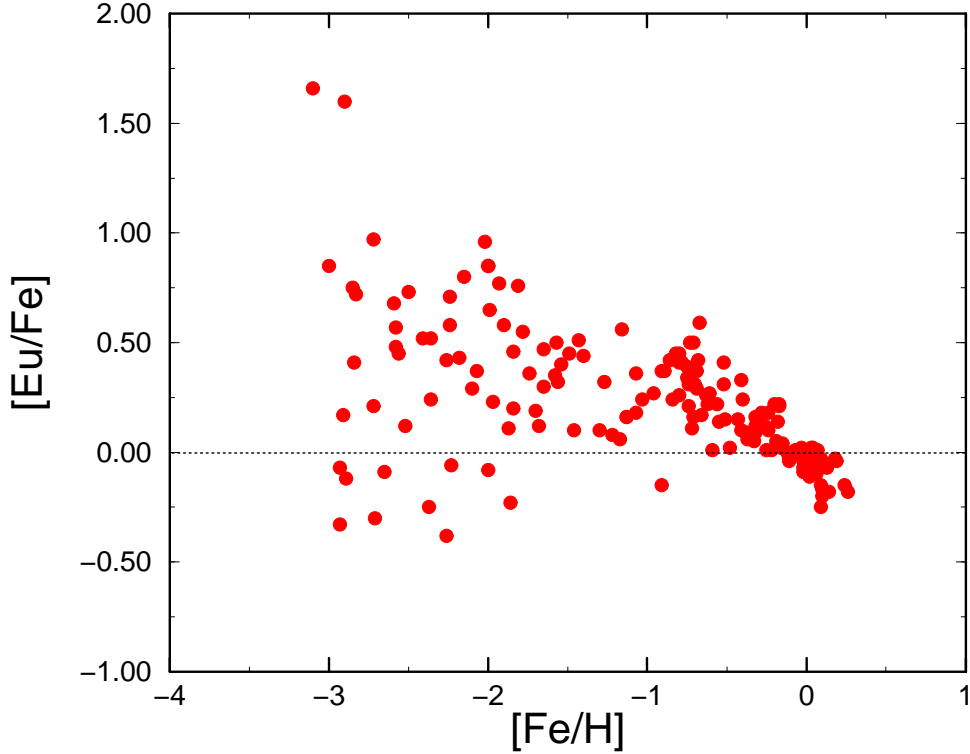}
\vskip 0pt \caption{
Abundance scatter of [Eu/Fe] versus metallicity for samples of halo and
disk stars, after Sneden \& Cowan (2003). 
}
\label{fig7.ps}
\end{figure*}

Further clues about the nature of the $r$-process are  found in 
examining the
abundance scatter of $n$-capture elements in the early Galaxy.
This trend was first noted by 
Gilroy et al. (1988) and then 
studied in more detail by Burris et al. (2000). 
We show in  Figure~\ref{fig7.ps}  [Eu/Fe] as a function of metallicity  
for a number of halo and disk stars from 
Sneden \& Cowan (2003) (see also Truran et al. 2002 and references therein).
The increasing level of star-to-star 
scatter of [Eu/Fe]  with decreasing metallicity,
particularly at values below [Fe/H] $\approx$ --2.0, 
suggests an early,
chemically unmixed and inhomogeneous Galaxy.
These data also suggest 
that not
all early stars are sites for the formation of both
$r$-process nuclei and
iron. 
Instead,  
this scatter is consistent with the
view that only a small fraction
(2\%--10\%)  of the massive stars that produce iron
also yield $r$-process elements (Truran et al. 2002).
Various  theoretical models to explain this abundance 
scatter have been proposed by,
for example,   Qian \& Wasserburg (2001) and
Fields, Truran, \& Cowan (2002).
Observations to help differentiate between these models and provide
new insight into the nature and site of $r$-process nucleosynthesis 
in the early Galaxy  are ongoing (Norris 2003).

\section{$r$-Process Chronometers}

The abundances  of  certain radioactive $n$-capture elements,
known as chronometers,  can be
utilized to obtain
age determinations  for the oldest stars, which in turn put lower limits
on age estimates for the Galaxy and the Universe.
Thorium, with a half-life of 14 Gy, in ratio to Nd (Butcher 1987) and to Eu 
(Pagel 1989) 
was  suggested as such a chronometer. 
Th/Eu is a preferred ratio---both are $r$-process elements and any 
possible  evolutionary effects of the predominantly $s$-process Nd 
are avoided. 
The detection of thorium in very metal-poor stars was pioneered by
Fran\c{c}ois, Spite, \& Spite (1993), and since then has been observed
in a number of these stars.
Chronometric ages, based upon the Th/Eu ratios, have
typically fallen in the range
of 11--15 Gyr for the observed stars
(e.g., Sneden et al. 1996;
Cowan et~al. 1997, 1999;
Pfeiffer, Kratz, \& Thielemann 1997; 
Sneden et al. 2000a, 2003; Westin et~al.
2000; Johnson \& Bolte 2001; 
Cowan et al. 2002).
Th/Eu ratios have also been determined for
several giants in the globular cluster M 15
(Sneden et~al. 2000b), who estimated their average, and hence the
cluster,  age at 14 $\pm$ 4 Gyr.

\begin{figure*}[t]
\includegraphics[width=1.00\columnwidth,angle=0,clip]{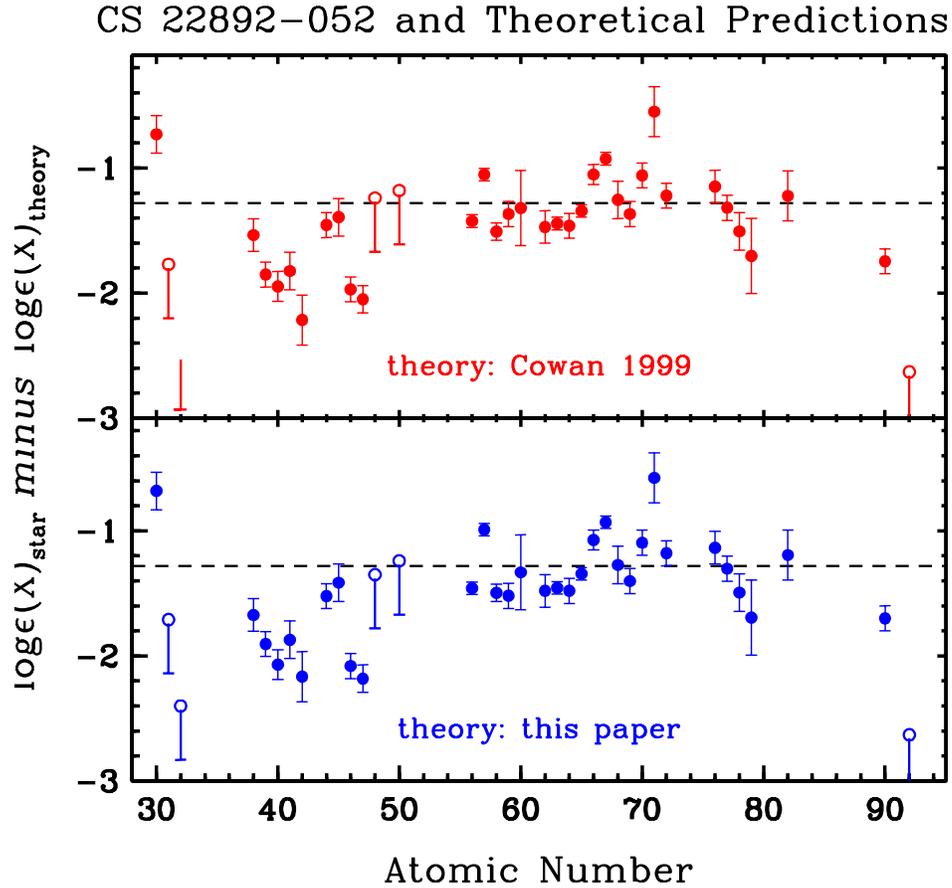}
\vskip 0pt \caption{
Differences between \cs22 abundances and two scaled $r$-process theoretical
predictions, after Sneden et al. (2003).  The top panel shows the abundance 
differences relative to those of Cowan \etal\ (1999), while the bottom 
panel differences are relative to Sneden et al.  (2003).
}
\label{fig8.ps}
\end{figure*}

These age estimates all typically have  errors 
$\sim \pm 3-4$ Gyr  
resulting from both observational and 
nuclear uncertainties.
In particular the chronometric age estimates
depend sensitively upon the initial predicted values of Th/Eu and 
hence on the nuclear mass formulae and $r$-process models
employed in making those
determinations.  We show, for example, in 
Figure~\ref{fig8.ps}
theoretical predictions for these ratios in comparison with recent
abundance determinations in \cs22 (Sneden et al. 2003). 
Utilizing the ETFSI-Q mass formula,  
the top panel shows predictions from Cowan et al. (1999),  while the bottom
panel shows a newer prediction, constrained by some recent experimental data
(Sneden et al. 2003).  These differences lead to age
uncertainties of $\sim$2 Gyr, 
while very different mass formulae lead to a wider range of
initial abundance ratios and correspondingly wider range in age estimates
(see Cowan et al. 1999; Truran et al. 2002).
The large  separation in nuclear mass number between Th and Eu might
also exacerbate uncertainties in these initial predictions
(see, e.g., Goriely \& Arnould 2001).
Thus,  it would be preferable to  
obtain abundances of stable elements nearer in mass number
to thorium (third $r$-process peak elements, for example), 
or, even better, to obtain two long-lived chronometers such as Th and U.

Uranium was first detected in any halo star by Cayrel et al. (2001),
who initially estimated the age of CS~31082$-$001 at 12.5 $\pm$ 3 Gyr. 
More refined abundance determinations (Hill et al. 2002) and 
recent theoretical studies have suggested an age of 15.5 $\pm$ 3.2 Gyr
(Schatz et al. 2002).
We note a fundamental difference in the abundance pattern of this 
star in comparison with several other metal-poor halo 
stars shown in 
Figure~\ref{fig4.ps}. Both Th and U are enhanced greatly with respect
to the other stable elements, including Eu, in CS~31082$-$001.
Thus, employing Th/Eu as a chronometer gives an unrealistic,
even a negative, age in this case
(Hill et al. 2002; Schatz et al. 2002;
see also Cayrel 2003 for a more complete discussion  of chronometers
in this star).
Interestingly,  
another tentative U detection has been recently
reported in the  metal-poor halo star \bd17 
(see Fig.~\ref{fig4.ps}), and in this case,
Th/U and Th/Eu give comparable age values.
It is not clear yet why CS~31082$-$001, with its
very large overabundances of Th and U, is so different,
but it clearly suggests that Th/U is a more reliable chronometer 
than Th/Eu---certainly it is in  this star. 
Unfortunately, it may be difficult to 
detect uranium in many halo stars; note the nondetection of this
element in \cs22 (Sneden et al. 2003).
Additional such detections of U, as well as  
continually improving nuclear descriptions of the very neutron-rich 
nuclei participating in the $r$-process, will be necessary to strengthen
this technique and reduce the age uncertainties.

\section{Summary and Conclusions}

A wealth of stellar abundance data has been assembled during the last
several decades. These high-resolution spectroscopic studies of 
low-metallicity halo stars---accompanied by significant new
experimental atomic physics data---have provided significant  clues about the
nature of the $r$-process and, at the same time, have imposed 
strong constraints on astrophysical  and nuclear model calculations.
The heavy $n$-capture abundances, from Z $\ge$ 56, 
appear to be consistent
with the relative solar system $r$-process abundance fractions, at 
least for the $r$-process-rich stars. This consistency suggests a 
similar mechanism, or well-constrained astrophysical 
conditions, for the operation of 
the $r$-process over many billions of years.
While there are less data available of the lighter $n$-capture elements,
the abundances of those elements appear to be not consistent---on average 
lower---with 
the same scaled solar system $r$-process distribution that fits the 
heavier $n-$capture elements. 
Various explanations have been offered to explain this difference
between the lighter and heavier $n$-capture 
abundance distributions: the possibility of two astrophysical sites
for the $r$-process (e.g., different masses or frequencies of supernovae,
or a combination of supernovae and neutron star binary mergers), or 
models with different conditions in the same
single core-collapse supernova. At this time, however,
it is not clear what the exact causes are for the apparent differences in
the abundance distributions of the lighter and heavier $n$-capture elements.

While the actual site for the $r$-process 
has not been definitively identified, 
there have been many advances in our understanding of the astrophysical
models and the related nuclear physics of this nucleosynthesis.
Core-collapse supernovae remain a promising site for the origin of the
$r$-process nuclei.
Much of the recent focus has been on obtaining more physically reliable
supernova models (e.g., including an improved treatment of neutrino processes).
Neutron star binary mergers,  with improved treatments of their evolution
and coalescence,  
have also been studied and can still be
considered a possible site for the $r$-process.
Accompanying these improved astrophysical models has been more 
experimental nuclear data and more reliable theoretical
prescriptions for neutron-rich nuclei far from stability.
 
The abundance patterns in the oldest Galactic halo stars  
have also provided additional new insights into the origins  and sites of
the $r$-process.  First, the elemental and isotopic abundances in the oldest
halo stars are consistent with an $r$-process-only 
origin at the earliest times in the history of the Galaxy.
These results suggest that the $r$-process sites in the earliest
stellar generations, the progenitors of the halo stars, were rapidly 
evolving---ejecting $r$-process-rich material into the interstellar medium long before
the major onset of Galactic $s$-process nucleosynthesis from low- 
and intermediate-mass stars.
In addition,  the star-to-star 
abundance scatter (e.g., [Eu/Fe]) observed in the lowest
metallicity (i.e., oldest) Galactic halo stars places strong constraints
on models of nucleosynthesis and  suggests that not all early stars were  
sites for the formation of both $r$-process nuclei and iron.
Further,  this abundance scatter  suggests an early chemically unmixed Galaxy.

The detection of the long-lived chronometers 
thorium, and now uranium, in some of the metal-poor halo
stars has allowed for the radioactive dating of the oldest stars.
This technique depends sensitively upon the observed stellar values and 
the theoretical predictions of
the initial abundance ratios (Th/Eu, Th/U, etc.) of  
elements synthesized in the $r$-process.
While there have been significant advances in nuclear physics,
both in experiment and theory, we still need to 
better define the properties of very $n$-rich (radioactive) nuclei.
These improvements  will be necessary  to 
better understand the origin and nature of the $r$-process and to  
reduce chronometric age uncertainties---strengthening 
the radioactive dating technique, and
providing more
precise age estimates for the Galaxy and the Universe.

\vspace{0.3cm}
{\bf Acknowledgements}.
We  thank all of our  colleagues for their contributions to the work described 
here.  We also thank the referee, Friedel Thielemann, for helpful comments.
This research has been supported in part by NSF grants AST-9986974  and 
AST-0307279 (JJC), AST-9987162  and AST-0307495 (CS), and by STScI grants 
GO-8111 and GO-08342.

\end{document}